\documentclass[9pt, twocolumn, conference]{IEEEtran}
\IEEEoverridecommandlockouts  
\usepackage{cite}
\usepackage[section]{placeins}
\usepackage{graphicx,subfigure}
\usepackage{amsmath,amssymb}
\usepackage{url}
\usepackage{color}
\usepackage{float}
\usepackage{flushend} 
\usepackage{dsfont}
\usepackage{setspace}
\usepackage{epstopdf}
\usepackage{algorithm}
\usepackage{algpseudocode}
\usepackage{caption}
\usepackage{multirow}
\usepackage{booktabs}
\usepackage[round, sort]{natbib}

\newcommand{\bx}{\boldsymbol{x}}

\newcommand{\ba}{\boldsymbol{a}}
\newcommand{\bb}{\boldsymbol{b}}

\newcommand{\bA}{\boldsymbol{A}}
\newcommand{\bB}{\boldsymbol{B}}

\newcommand{\bX}{\boldsymbol{X}}

\begin{document}

\allowdisplaybreaks
\title{\huge Energy Disaggregation with Semi-supervised Sparse Coding}
\author{Mengheng~Xue, Samantha Kappagoda and David K. A. Mordecai\\
        RiskEcon\textsuperscript{\textregistered}  Lab for Decision Metrics @ Courant Institute of Mathematical Sciences\\
        New York University, New York, NY 10011\\
        mx586@nyu.edu, \{kappagoda, mordecai\}@cims.nyu.edu
}


\maketitle
\thispagestyle{plain}
\pagestyle{plain}
\allowdisplaybreaks
 
\begin{abstract}
Residential smart meters have been widely installed in urban houses nationwide to provide efficient and responsive monitoring and billing for consumers. Studies have shown that providing customers with device-level usage information can lead consumers to economize significant amounts of energy, while modern smart meters can only provide informative whole-home data with low resolution. Thus, energy disaggregation research which aims to decompose the aggregated energy consumption data into its component appliances has attracted broad attention. In this paper, a discriminative disaggregation model based on sparse coding has been evaluated on large-scale household power usage dataset for energy conservation. We utilize a structured prediction model for providing discriminative sparse coding training, accordingly, maximizing the energy disaggregation performance. Designing such large scale disaggregation task is investigated analytically, and examined in the real-world smart meter dataset compared with benchmark models. 

\end{abstract}

\section{Introduction}
Energy efficiency has raised increasing public concerns due to the persistent and predominant consumption of fossil fuels, which has remained largely unabated during recent decades \citep{Mobilia2017even}. More recently, smart meters and other metering sensors interconnected across distributed networks are being increasingly adopted in urban households \citep{benzi2011electricity},  playing crucial roles in providing efficient, secure and reliable residential energy usage monitoring \citep{zheng2013smart, krishnamurti2012preparing}. As of 2018, more than 86.8 million smart meters have been installed across the US, among which 88\% are equipped in residential houses \citep{chui2018energy}. Smart meters record energy at a low resolution (typically every hour or 15 minutes) and reported at least daily \citep{feuerriegel2018more}. Although residential smarter meters can report detailed information on whole-home electricity consumption in real-time, they do not perform appliance-level monitoring of power usage, which would provide more informative feedback for customers to conserve energy. According to \cite{darby2006effectiveness}, with detailed appliance-level energy consumption notification,  residents can modify their consumption habits, which can improve energy efficiency by 12\%. What's more,  disaggregated energy information plays a crucial role in diagnosing household electricity problems and forecasting demand \citep{froehlich2010disaggregated}. Therefore, energy disaggregation (also referred to as non-intrusive load monitoring or NILM), which aims to estimate the power demand of individual appliances based on a single whole-home aggregated energy signal, has attracted great interest in the applied machine learning area \citep{herrero2017non}. 

NILM typically monitors a single source of usage data by employing signal processing to identify constituent components based upon a data repository (i.e., a {\em dictionary}) of current and voltage signatures. Aside from electricity usage analytics, current and prospective use-case applications also include condition-based maintenance and anomaly detection related to intrusions, faults and defects (e.g., short circuits, overloads), in order to prevent or mitigate the risk of outages and electrical fires. In the disaggregation and recognition of energy consumption patterns, sampled signals from an electricity consumption indicator are disaggregated in order to identify device and appliance level power usage patterns, often by employing machine learning or other signal processing methods. The notion of energy usage disaggregation underlying NILM is straightforward: monitor a single source of data on energy usage (e.g., a smart meter) and employ signal processing software to identify constituent components based upon unique energy-usage (e.g., current and voltage) signatures. The objective of such signal disaggregation is to identify energy consumption patterns at a  much lower cost than would be incurred with submetering. 

By way of example, one such approach employs two {\em current transformers} (CTs) as sensors that measure alternating current (AC) attached to the service mains of the electric panel for a property, in order to measure electricity consumption (or generation). These current sensors monitor usage patterns, in order to disaggregate the consumption signatures of individual appliances by sampling current and voltage at a rate of one million times per second. This high-resolution data is used to train algorithms in order to detect minute variations in magnitude, phase, and frequency to differentiate between appliance performance. Split-core type CTs have a primary winding, a magnetic core, and a secondary winding and may be attached as sensors to either live or neutral wiring entering a building, without the need for high voltage electrical work. For whole building monitoring, the primary winding is the live or neutral wire (but not both) entering the building, passed through the opening in the CT with the secondary winding consisting of many turns of fine wire housed within the transformer case.

\subsection{Existing Research}  
As surveyed by \cite{butzbaugh2019residential}, the literature is extensive regarding the suitability, reliability, robustness and accuracy of particular algorithms and sensors for NILM and disaggregation of energy usage patterns at device and appliance levels.  It  has  been further noted  throughout the literature that  available  sensing devices  on  the  market  tend  to  be  highly  variable  in  relative  accuracy for identifying usage patterns. Studies assessing multiple disaggregation tools document that the accuracy of a given product in identifying the energy consumption associated with a particular end use (e.g., a refrigerator) can range between 0 to 90 percent.  Energy disaggregation research introduced in \citep{hart1992nonintrusive} can be classified as supervised and unsupervised methods depending on whether a training dataset of power consumption from individual appliances is utilized. For supervised algorithms, optimization and pattern recognition techniques are commonly applied. Integer programming to determine the optimal combination of active appliance states is discussed in  \citep{suzuki2008nonintrusive}. The study \citep{baranski2004genetic} presents a genetic algorithm combining with fuzzy clustering to detect usage patterns using only standard digital meters. To detect power signals of devices, a multi-label classification framework employing both time and wavelet domain features is introduced in \citep{tabatabaei2016toward}. \cite{altrabalsi2016low} propose a k-means clustering SVM model to handle low-complex NILM problems with a short training period. Besides, deep neural nets have driven remarkable improvements in the NILM area. In \citep{kelly2015neural}, both convolutional (CNN) and recurrent neural networks (RNN) are employed to model sequence-to-sequence (seq2seq) learning. The study \citep{bonfigli2018denoising} examines a denoising autoencoder which attempts to reconstruct the clean power demand of target appliances.  On the other hand, unsupervised work utilizes the Factorial Hidden Markov Model (FHMM) considering additional features such as time of day and dependency between appliances is described in \citep{kim2011unsupervised} to infer the state of each appliance.  Following that, \cite{kong2016hierarchical} consider a hierarchical extension of the HMM model to handle the appliances with multiple functional modes. 

\subsection{Contributions}
Most existing studies within both supervised and unsupervised frameworks have their limitations. Supervised learning requires high-quality individual appliance data for model training, before the system deployment, in order to guarantee accurate disaggregation results. Although unsupervised approaches do not rely on the preceding appliance-level data,  most are highly task-dependent \citep{kim2011unsupervised} and their disaggregation results usually underperform the supervised models with extra training stages. Additionally, most studies are conducted in a laboratory environment containing household energy data within a relatively small number of houses, e.g., REDD \citep{kolter2011redd} and BLUED \citep{filip2011blued}. Therefore, the generalizability of such algorithms has not been fully investigated yet. 

Motivated by a more practical strategy to conduct energy disaggregation, we adopt the semi-supervised approach based on sparse coding to disaggregate power usage using low-resolution, hourly data which is readily accessible via smart meters. Specifically, we apply the {\em discriminative} training algorithm based on sparse coding to learn hidden patterns (i.e.,  {\em sparse representations}) in each appliance's usage over a typical week, then combine with the learned hidden patterns to predict the appliance-level power usage of unseen houses using their whole-home aggregate signal alone. Furthermore, we consider a substantially larger household energy dataset containing approximately 1,000 houses within a two-year period  and conduct experiments using selected quantitative metrics to evaluate the algorithm efficiency in realistic scenarios.


\section{System Model}
\label{sec:System Model}

\subsection{Data Model}
\label{sec:data}
Consider a lossless household power system consisting of $M$ households and $K$ number of appliances, e.g., refrigerators and heaters,  with $T$ hours power usage recording  in each household. We denote the energy usage recorded for device $k \in \{1, \ldots, K\}$ in household $m \in \{1,\ldots M\}$ during $T$ hours with $\bx^m_k\in \mathbb{R}^{T\times 1}$. Accordingly, we represent energy usage recording for a particular device as one class:
\begin{equation}
\bX_k \triangleq \left[{\bx^{1}_k}, \ldots, {\bx^{M} _k}\right]\;, 
\end{equation}
where $\bX_k\in\mathbb{R}^{T\times M}$. Based on this, we define the {\em aggregated} power consumption including all  $K$ devices as 
\begin{equation}
\bar{\bX}\triangleq \sum_{k=1}^K \bX_k\;, 
\end{equation}
where $\bar{\bX}\in\mathbb{R}^{T\times M}$ and the $m^{\rm th}$ column of $\bar{\bX}$ contains $T$ hours aggregated energy consumption for all devices in a given house $m$. At the training stage, we  assume that the individual device energy readings $\bX_1, \ldots, \bX_K$ are accessible from plug-level or outlet-level monitors in a limited number of instrumented homes. While at the test stage, we assume that we can only access the aggregated signal $\bar{\bX}'$ reported by smart meters, and we aim to separate the aggregated signal into its component appliances, i.e., $\bX_1', \ldots, \bX_K'$. 

\subsection{Non-negative Sparse Coding Model}
\label{sec:NSC}

Sparse coding has been widely applied as an effective data representation method to source separation problems \citep{schmidt2006single, schmidt2007wind}. We adopt this approach to train separate models for each individual class $\bX_k$, then apply these models to separate aggregated signals in the testing stage. Specifically,  sparse coding attempts to approximate each class training sample $\bX_k$ as 
\begin{equation}
\bX_k \approx \bB_k\bA_k\;,
\end{equation}
where $\bB_k = [\bb_1, \ldots, \bb_n]\in \mathbb{R}^{T\times n}$ represents a {\em dictionary} matrix and its columns are $n$ basis functions, and $\bA_k = [\ba_1, \ldots, \ba_M]\in \mathbb{R}^{n\times M}$ represents a {\em activation} matrix and each column represents the activation of basis functions \citep{wang2014semi}.  It is noticeable that sparse coding assumes that only a few basis functions in the dictionary are sufficient to represent the training data, thus the activation matrix $\bA_k$ should be sparse, i.e., most of the elements are zeros, leaving only a few non-zeros. Accordingly, an {\em over-complete} dictionary of the training data in which more basis functions than the dimensions of training data (i.e., $n \gg T, M$) is learned. To solve the sparsity, we need to impose an $\ell_1$ regularization penalty to the activations \citep{lee2007efficient}. 

What's more, since the input data (energy usage) is inherently non-negative, it imposes the further constraints that the dictionary and activation matrices are both non-negative. Accordingly, the non-negative sparse coding (NNSC) for energy usage is to find non-negative optimal sparse representation of training data such that the reconstitution error is minimized:
\begin{equation}\label{nnsc}
\begin{aligned}
& \underset{\bA_k \geq 0, \bB_k \geq 0}{\text{minimize}}
& & \frac{1}{2} \|\bX_k-\bB_k\bA_k\|^2_F+\lambda\|\bA_k\|^2_F\\
& \text{subject to}
& & \|\bb_k^{(j)}\|_2 \leq 1, \; \forall j \in \{1,\dots, n\}\\
\end{aligned}\;,
\end{equation}
where $\|\cdot\|_F$ is the Frobenius norm,  $\|\cdot\|_2$ is the $\ell_2$ norm and $\lambda\in \mathbb{R}^+$ is a regularization parameter to encourage sparsity of activation matrix $\bA_k$. Moreover, $ \|\bb_k^{(j)}\|_2 \leq 1$ constraints are imposed to  reduce the complexity of each basis function. It is noticeable that \eqref{nnsc} is not jointly convex in $\bA_k$ and $\bB_k$, but convex in each variable when holding the other fixed. Thus, to minimize the objective with respect to both, we can take turns updating $\bB_k$ and $\bA_k$  \citep{hoyer2002non}.

Based on the above procedure, sparse representations $\bA_k$ and $\bB_k$ for power usage $\bX_k$ of each class $k \in \{1, \ldots, K\}$ in the training dataset can be learned based on \eqref{nnsc}. Since $\bB_k$ is trained to reconstruct the $k{\rm th}$ class with small activation, we assume it is the optimal dictionary to reconstruct the $k{\rm th}$ portion of the aggregated signal.  In the testing stage, we assume that the bases learned from the training stage remain the same for reconstructing signals, i.e., 
\begin{equation}\label{ass1}
\bB_k' = \bB_k, \forall k \in \{1, \ldots, K\}\;.
\end{equation}
We aim to disaggregate a new aggregated power signal $\bar{\bX}'\in \mathbb{R}^{T\times M}$ without providing any information on its components.  When the sparse coding framework is applied to the test aggregated signal and we assume that the sources are additive, we have
\begin{equation}\label{testsum}
\bar{\bX}'  = \sum^K_{k=1} \bX_k' \approx \left[\bB_1, \ldots, \bB_K\right]  \begin{bmatrix} \bA_1'\\
\vdots \\
\bA_K'
\end{bmatrix} \triangleq \bB_{1:K} \bA_{1:K}'^T\;,
\end{equation}
where we denote $\bB_{1:K}$ as shorthand for $[\bB_1, \ldots, \bB_K]$.  Therefore, based on \eqref{testsum}, the joint set of optimal new activation matrices $\hat{\bA'}_{1:K}$ can be computed  as the following problem:
\begin{equation}\label{newac}
\boldsymbol{\hat{A}}'_{1:K} = \text{arg} \underset{\bA'_{1:K} \geq 0}{\text{min}} \|\bar{\bX}' - \bB_{1:K}(\bA_{1:K}'^T)^T\|_F^2 + \lambda\|\bA'_{1:K}\|^2_F\;. 
\end{equation}
Therefore, based on the basis $\bB_k$ learned from training stage and the activation $\boldsymbol{\hat{A}}'_k$ computed in \eqref{newac}, we can predict the $k{\rm th}$ component of tested aggregated signal $\bar{\bX}'$ as 
\begin{equation}
\hat{\bX}'_k = \bB_k \boldsymbol{\hat{A}}'_k, \quad\forall k\in \{1, \ldots, K\}\;. 
\end{equation} 
Therefore, the {\em disaggregation error} is defined as 
\begin{equation}\label{err}
\begin{aligned}
& E \triangleq\sum^K_{k=1}\frac{1}{2} \|\bX_k - \bB_k\boldsymbol{\hat{A}}_k\|^2_F\\
& \text{subject to}\\
& \boldsymbol{\hat{A}}_{1:K} = \text{arg} \underset{\bA'_{1:K} \geq 0}{\text{min}}\left\|\bar{\bX}- \bB_{1:K}(\bA_{1:K}^T)^T\right\|_F^2 + \lambda\|\bA_{1:K}\|^2_F
\end{aligned},
\end{equation}
which evaluates the accuracy for reconstructing individual classes via only the aggregated signal combining with obtained activations.

\subsection{Discriminative Disaggregation Model}
The sparse coding approach only concerns the sparse approximation of input data, and it strongly hinges on the assumption that the disaggregation dictionary is the same as that of reconstruction in \eqref{ass1}. However, the additional discriminative information provided by bases over the training signals is limited when being used in complex disaggregation tasks \citep{wang2012signal}. Accordingly, {\em discriminative disaggregation sparse coding} (DDSC) which is to learn a dictionary whose resultant activations possess improved discriminative power is proposed in  \citep{kolter2010energy}. Specifically, with the assumption that the disaggregation dictionary is not necessarily the same as that for reconstruction, and the intuition that optimal value of $\boldsymbol{\hat{A}}_k$ should be the activations learned by sparse coding is given by $\forall k \in \{1, \ldots, K\}$, 
\begin{equation}\label{a_star}
\bA^*_k  = \text{arg} \underset{\bA_k \geq 0}{\text{min}} \frac{1}{2} \|\bX_k-\bB_k\bA_k\|^2_F+\lambda\|\bA_k\|^2_F\;, \\
\end{equation}
DDSC aims to minimize disaggregation error in \eqref{err} while {\em discriminatively} optimizing the the dictionary bases in order to move activations $\hat{\bA'}_{1:K}$ as close to $\bA^*_{1:K}$ as possible \citep{wang2012signal}. Therefore, the disaggregated error in \eqref{err} can be converted into an {\em augmented regularized disaggregation error} defined as

\begin{equation}\label{aug_err}
\begin{aligned}
& \tilde{E} \triangleq \sum^K_{k=1}\left( \frac{1}{2} \|\bX_k - \bB_k\boldsymbol{\hat{A}}_k\|^2_F+ \lambda\|\boldsymbol{\hat{A}}_k\|^2_F\right)\\
& \text{subject to}\\
& \boldsymbol{\hat{A}}_{1:K} = \text{arg} \underset{\bA_{1:K} \geq 0}{\text{min}} \left\|\bar{\bX}- \tilde{\bB}_{1:K}(\bA_{1:K}^T)^T\right\|_F^2 + \lambda\|\bA_{1:K}\|^2_F
\end{aligned},
\end{equation}
where $\bB_{1:K}$ denotes the {\em reconstruction bases} which are learned from sparse coding in \eqref{nnsc}, and $\tilde{\bB}_{1:K}$ denotes the {\em discriminative bases} which intend to reduce the difference between $\boldsymbol{\hat{A}}_{1:K}$ and $\bA^{*}_{1:K}$ where $\bA^{*}_{1:K}$ is the activations learned by sparse coding in \eqref{a_star}. To update the discriminative bases $\tilde{\bB}_{1:K}$ and by denoting $\tilde{\bB} = \left[\tilde{\bB}_1,\ldots, \tilde{\bB}_K \right]$ and $\bA^*=\left[\bA^{*T}_1,\ldots, \bA^{*T}_K\right]^T$ (and similarly for $\boldsymbol{\hat{A}}$), we apply a structured perceptron algorithm \citep{collins2002discriminative} stated as

\begin{equation}\label{db}
\tilde{\bB} \leftarrow \tilde{\bB}-\alpha\left((\bar{\bX} - \tilde{\bB}\boldsymbol{\hat{A}})\boldsymbol{\hat{A}}^T - (\bar{\bX}-\tilde{\bB} \bA^*)\bA^{*T}\right)\;,
\end{equation}
where $\alpha$ denotes the perceptron updating step size. It is noticeable that for each iteration we need to remove negative elements and re-normalize each column in $\tilde{\bB}_{1:K}$ to keep its form consistent with $\bB_{1:K}$.

\subsection{Prediction Model}
\label{sec:prediction}

Based on the discriminative bases obtained in \eqref{db}, the {discriminative activations} $\tilde{\bA}'_{1:K}$ can be computed by reformulating \eqref{newac} into 

\begin{equation}\label{ak_star}
\tilde{\bA}'_{1:K}  = \text{arg} \underset{\bA'_{1:K} \geq 0}{\text{min}} \|\bar{\bX}' - \tilde{\bB}_{1:K}(\bA_{1:K}'^T)^T\|_F^2 + \lambda\|\bA'_{1:K}\|^2_F\;.
\end{equation}
Accordingly, the prediction of the $k$th component of new aggregated signal $\bar{\bX}'$ based on DDSC can be expressed as 
\begin{equation}
\tilde{\bX}_k' = \tilde{\bB}_k \tilde{\bA}'_k, \quad\forall k\in \{1, \ldots, K\}\;. 
\end{equation}
Therefore, we summarize the iterative DDSC algorithm as follows: the NNSC algorithm is employed first to initialize the target activations and reconstruction bases. Following that, the discriminative disaggregation (DD) training iterations in \eqref{aug_err} and \eqref{db}  are repeated until convergence and the discriminative bases are obtained. Finally, given the aggregated test examples and the estimated activations, the disaggregated solutions for the test examples are outputted. The detailed pseudo-code for the DDSC algorithm is presented in \citep{kolter2010energy}.

\section{Experimental Results}
\label{sec:results}
In this section, we evaluate the performance of the DDSC algorithm on real-world scenarios. 
\subsection{Dataset and Experimental Setup}\label{subsec:data}
We conduct experiments using the Pecan Street dataset, which contains 60-second interval energy readings from nearly 1,000 houses with up to 23 domestic appliance-level meters per household, collected over two years across three research sites (i.e., Austin, New York, California) \citep{street2015dataport, russo2019enabling}.  
Since the widely deployed smart meters are constrained to report at a low sample rate (typically $1/3600$ Hz), for generalization purposes, we convert both whole-home and device-level readings into interval-based time series with a $1/3600$ Hz sample rate \citep{liang2019hvac}. Meanwhile, we reduce the size of bases to guarantee the performance of DDSC by only considering five categories of electrical devices with distinct consumption patterns (i.e., furnace, dishwasher, refrigerator, air, and a miscellaneous category) \citep{dong2013deep}. Due to limited computational power, we choose to select 80 houses that satisfy the requirements as our dataset, i.e., $m = 80$. Critically, we focus on disaggregating data from homes that are absent from the training data, where we assign $70\%$ of the data for training and $30\%$ of the data for testing. The hyper-parameters of the algorithm (e.g., number of bases $n$, regularization parameter $\lambda$ and step size $\alpha$) are all selected independently using grid search through a discrete set of empirical values. Again, due to the computational limits, we provide the weekly ($T=168$) disaggregated predictions in different cases to evaluate the DDSC performance. 

\subsection{Implementation}
Based on the dataset and settings described in Section \ref{subsec:data}, we conduct all DDSC experiments using Python 3.7 packages  \citep{oliphant2007python}  including machine learning solver Scikit-learn \citep{pedregosa2011scikit} and signal processing solver Librosa \citep{mcfee2015librosa}. Specifically, we implement a coordinate decent approach \citep{friedman2007pathwise} to solve the optimization problems in \eqref{a_star} and \eqref{aug_err} using Scikit-learn module SparseCoder and  Librosa to decompose and retrieve the activation matrices. In the NNSC pre-training stage, 
to solve the optimization over $\bB_k$ in \eqref{nnsc}, we apply the multiplicative non-negative matrix factorization method proposed by   \cite{eggert2004sparse}. For the rest part, we rely on the standard methods for solving the  optimization problems. 

\subsection{Weekly Energy Disaggregation}

\begin{figure*}[t]
     \begin{center}
        \subfigure[Usage prediction in winter.]{
           \includegraphics[width=3.2in,height=4.45in]{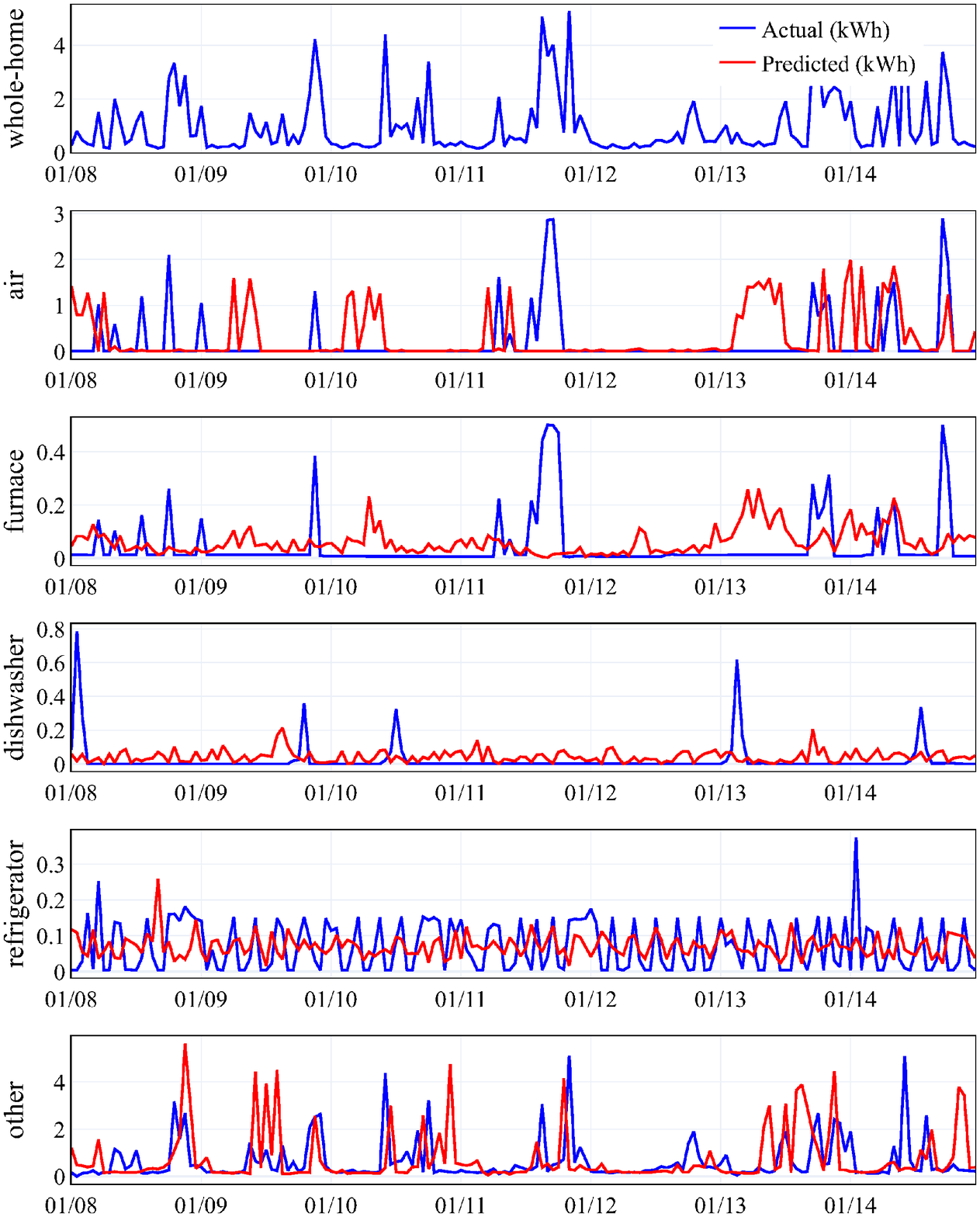}
            \label{winter}
        }\hspace{.4 in}
        \subfigure[Usage prediction in summer.]{
           \includegraphics[width=3.2in,height=4.45in]{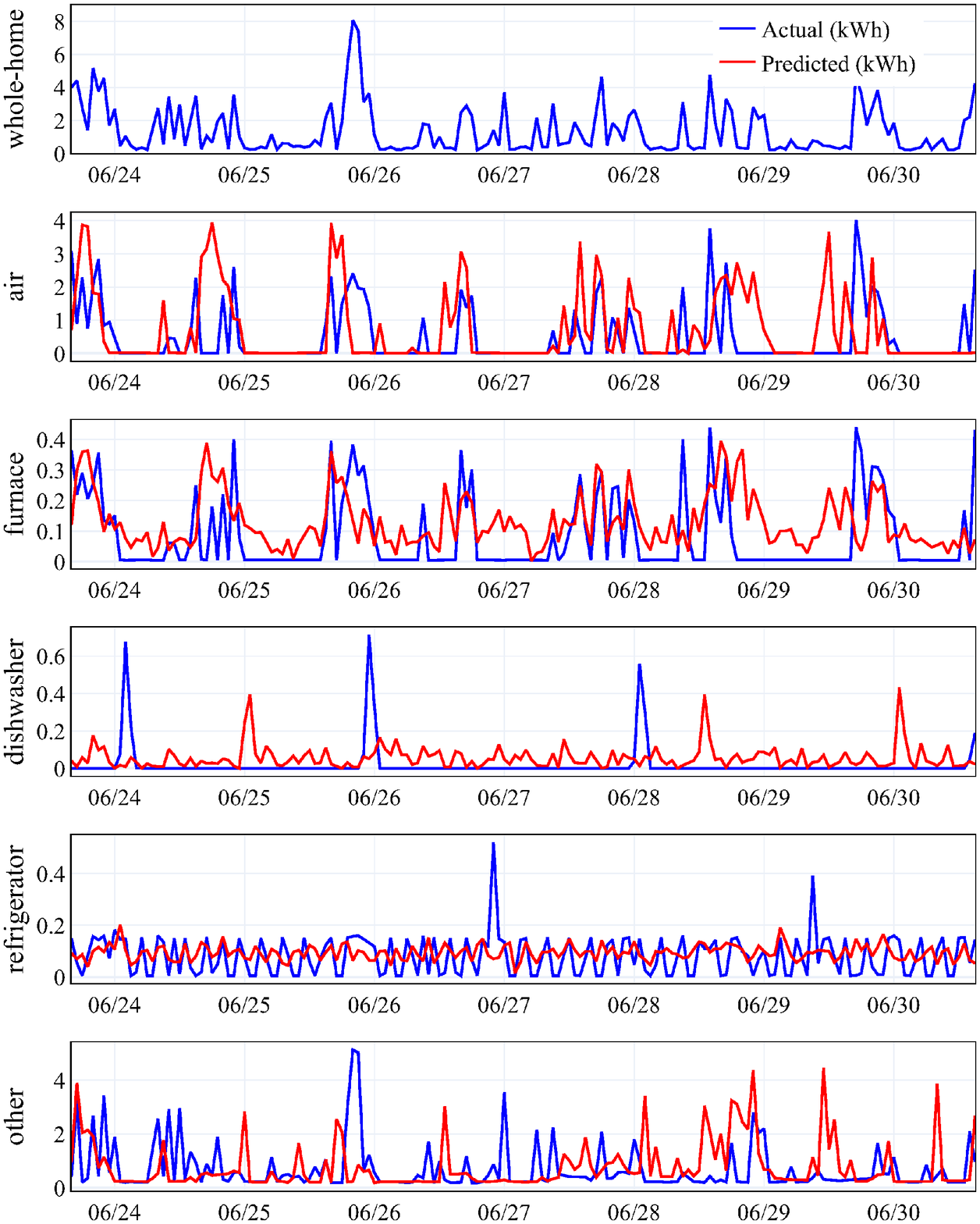}
            \label{summer}
        }\hspace{.4 in}
        \subfigure[Percentage prediction in winter.]{
           \label{winter_per}
           \includegraphics[width=3.2in,height=1.8in]{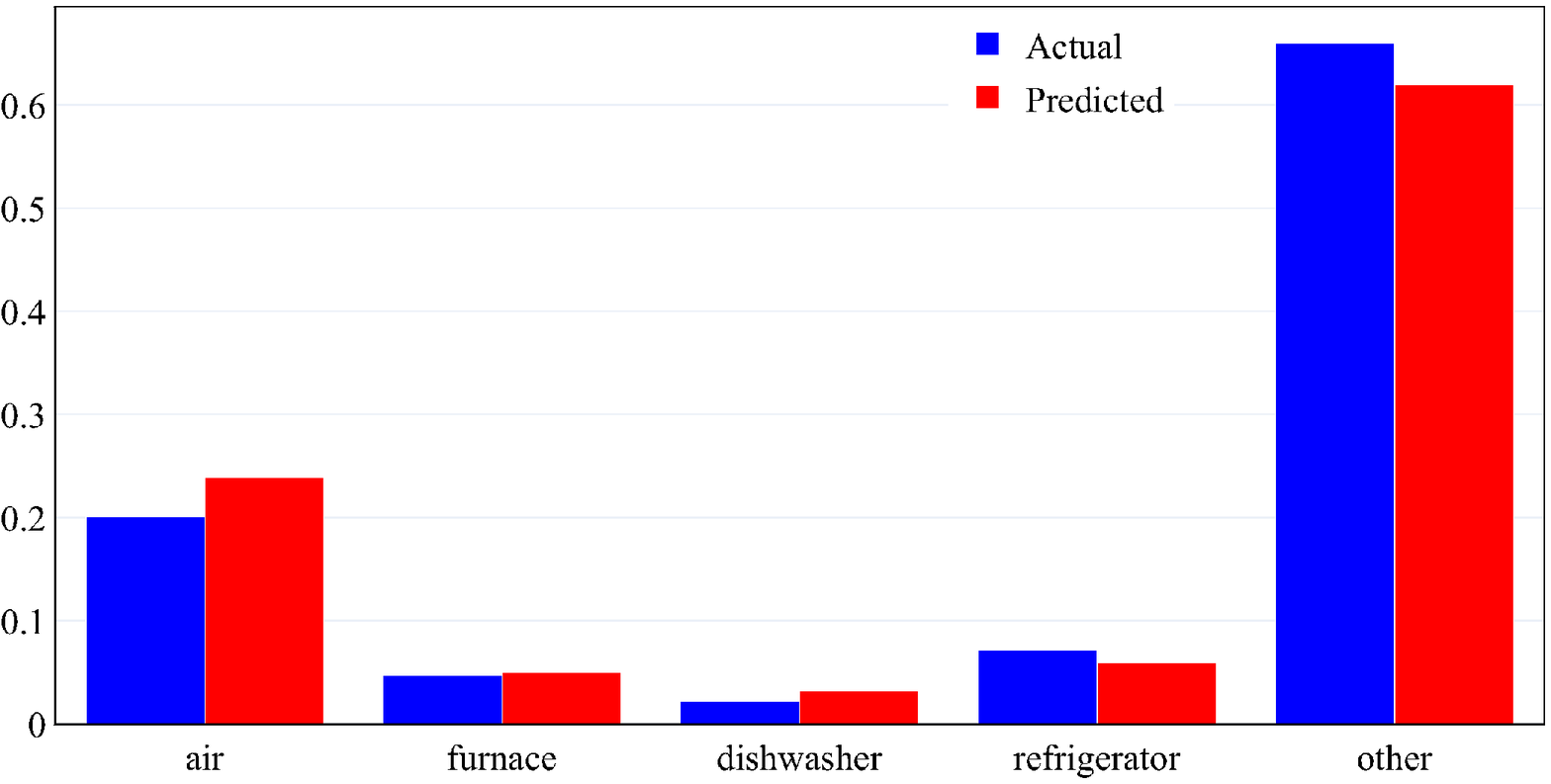}
       }\hspace{.4 in}
       \subfigure[Percentage prediction in summer.]{
           \includegraphics[width=3.2in,height=1.8in]{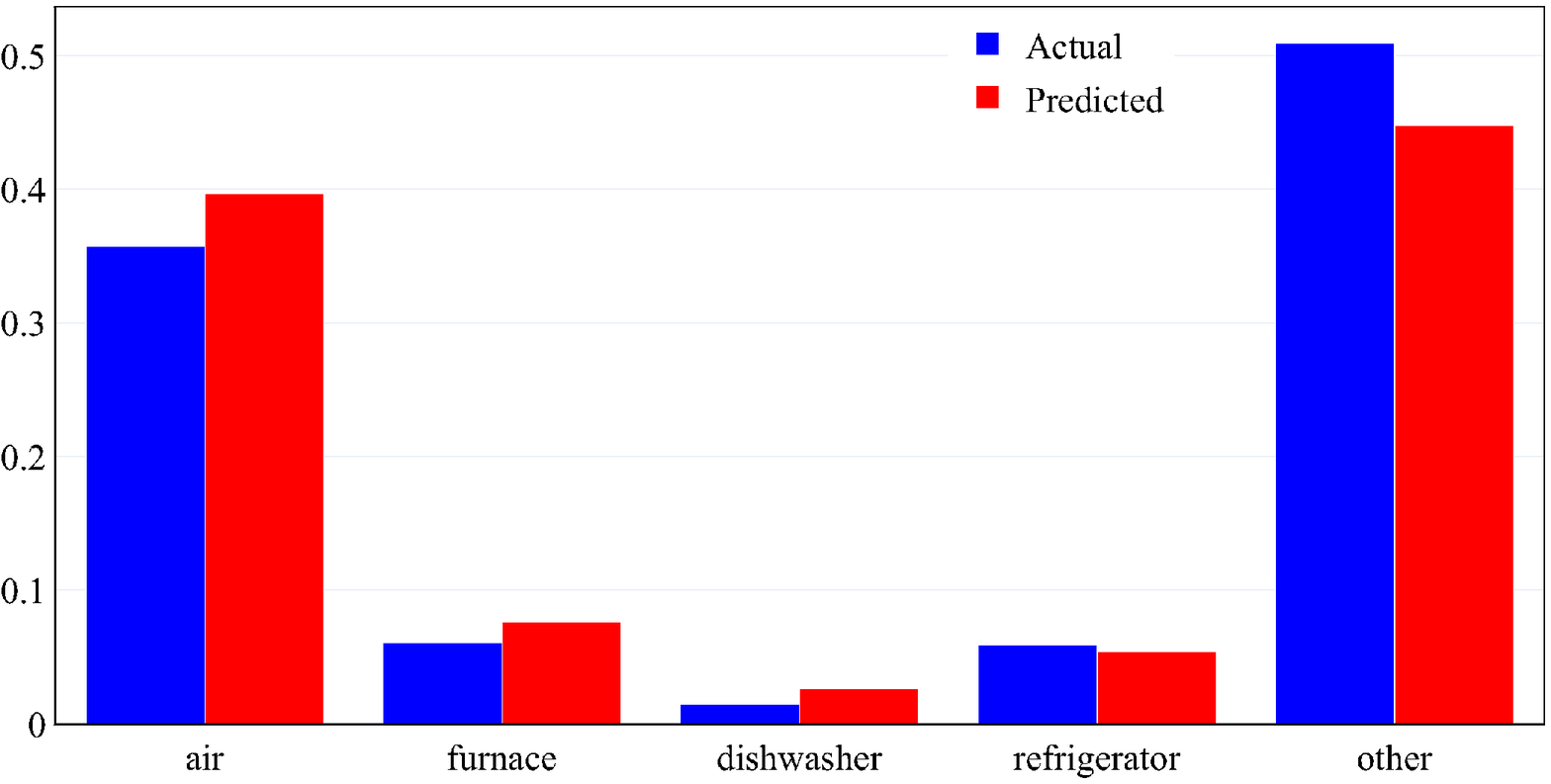}
            \label{summer_per}
        }\hspace{.4 in}
       	\caption{Example weekly predicted energy profiles and total energy percentage in winter and summer periods.}
       	\label{weekly}
    \end{center}
\end{figure*}

In this subsection, we evaluate the weekly prediction of energy disaggregation using DDSC during different seasons. Fig.~\ref{winter} and \ref{summer}  illustrate the true energy consumption of one selected house in the test set for two different weeks in summer and winter respectively, along with the predicted consumption by DDSC. Fig. ~\ref{winter_per} and ~\ref{summer_per} show the relative consumption comparison of different devices over the whole week in both seasons. 

We can see that for the selected house, weekly energy usage in the summer is higher than that in the winter mainly due to the relatively heavier usage of air conditioners. In contrast, the refrigerator has steady usage patterns regardless of seasons. In terms of the total predicted percentage,  we can see that DDSC performs well for this particular house in both seasons. In terms of usage signals, for devices which hold certain steady usage pattern, e.g., refrigerator, DDSC can capture the weekly trend well. However, DDSC fails to predict the spikes in dishwasher usage. The underlying reason is that when the component signal is too sparse and relatively weak, it is hard for DDSC to identify the disaggregation dictionary when training bases $\tilde{\bB}_{1:K}$  with limited label data \citep{dong2013deep}. Moreover, DDSC has difficulty in disaggregating air and furnace components during winter. This can possibly  be explained that the consumption patterns of air and furnace are highly correlated with each other, which leads to poorer disaggregation performance.


\subsection{Quantitative Evaluation}
Next, we present metric results that evaluate the performance of DDSC on each of these components compared with the baseline NNSC model. When we focus on the error in energy usage at every time point, we apply the mean absolute error (MAE) defined as
\begin{equation}
{\rm MAE} = \frac{1}{T}\sum^T_{t=1} |\tilde{\bx}_k(t)-\bx_k(t)|\;, \quad\forall k\in \{1, \ldots, K\}\;,
\end{equation}
where $\bx_k(t)$ denotes the ground truth and $\tilde{\bx}_k(t)$ denotes the prediction of device $k$ at time $t$. Meanwhile, when we are interested in recovering the total energy used by each appliance over a period, e.g., one week, we use normalized signal aggregated error (SAE) \citep{zhang2018sequence} expressed as 

\begin{equation}
{\rm SAE} = \frac{|\tilde{r}_k-r_k|}{r_k}\;, \quad\forall k\in \{1, \ldots, K\}\;,
\end{equation}
where $r_k$ and $\tilde{r}_k$ denote the ground truth and the predicted total power consumption of device $k$ over time period $T$, i.e., $\tilde{r}_k = \sum^T_{t=1}\tilde{\bx}_k(t)$ and $r_k = \sum^T_{t=1}\bx_k(t)$. This measure can provide accurate evaluation for weekly power usage even if its per-timestep prediction is less accurate. However, we do not apply the widely used Normalized Disaggregation Error (NDE) \citep{zhong2014signal} as 
\begin{equation}
{\rm NDE} = \frac{\sum^T_{t=1} \left(\tilde{\bx}_k(t)- \bx_k(t)\right)^2}{\sum^T_{t=1}\bx_k(t)^2}\;, \quad\forall k\in \{1, \ldots, K\}\;, 
\end{equation}
since it is less straightforward to provide meaningful feedback to householders compared with MAE. Also, MAE is more robust than NDE with outliers, i.e., extremely inaccurate predictions \citep{zhang2018sequence}.

\begin{table}[h!]
  \begin{center}
    \caption{Disaggregation result comparison between DDSC and NNSC based on MAE (kWh) and SAE  on Pecan Street dataset.}
    \label{tab:table1}
 \begin{tabular}{lcccl}\hline
& \multicolumn{2}{c}{MAE} & \multicolumn{2}{c}{SAE} \\
\cmidrule(lr){2-3}\cmidrule(lr){4-5}
Appliance & NNSC & DDSC & NNSC & DDSC \\
\hline
      Air &  1.031 & 0.732 & 1.286 & 0.413 \\
      Furnace &0.360 & 0.103 & 1.097& 0.527\\
      Dishwasher & 0.152&0.070  & 2.315& 1.741\\
      Refrigerator & 0.124& 0.067 & 0.563& 0.176\\
      Other & 1.656 & 0.733 &0.374 & 0.120\\
      Overall & 0.665 & 0.341 & 1.127&  0.595\\
      \hline
    \end{tabular}
  \end{center}
\end{table}

As we can see from Table~\ref{tab:table1}, DDSC outperforms NNSC on the Pecan Street dataset. Specifically,  DDSC reduces MAE by 48.7\% and SAE by 47.2\% overall compared with NNSC, which are also improved for each appliance. It clearly demonstrates that the discriminative training in DDSC is crucial to enhancing the performance of the energy disaggregation procedure and provides a significant improvement over the baseline NNSC model.

\section{Conclusion}
Energy disaggregation, which separates whole-home power consumption into individual device-level consumption, can make a significant contribution to energy conservation. In this paper, we examine the importance of structure learning for disaggregation tasks and present a sparse coding model to discover discriminative sparse structures of individual devices. The extensive experiments focusing on high volumes of disaggregated (and in many cases somewhat relatively low-resolution) data available from smart meters and related sensors may be conducted to demonstrate whether or not – and if so, under what conditions and to what degree – the discriminative model might substantially improve the performance of sparse coding in disaggregation tasks. Estimated consumption patterns being specifically inspected may provide detailed forensic measurement regarding energy consumption and other related decision support analytics for maintenance and risk mitigation.

\section{Discussion and Future Research}
The primary bottleneck in our approach is that the disaggregation performance is not reliable when training bases contain large amounts of appliances with similar consumption patterns and limited label metadata. Since the size of the dictionary matrix expands, the massive number of tunable parameters will lead to many local optima and high time consumption for iterative updates. Also, only a small amount of labeled data can be collected for training due to user privacy and security concerns. Therefore, a possible research direction is to combine sparse coding with deep learning to learn multiple layers of dictionaries to capture more nuanced hidden patterns of each device. 

We also plan to analyze the sensitivity of energy disaggregation performance to smart meter sampling frequency. More large-scale datasets and published algorithms will be incorporated into comprehensive comparison. Furthermore, some hybrid approaches that vary the sampling rate in order to track high-frequency patterns will be evaluated. Lastly, the applicability of the discriminative sparse coding method to other source separation problems will be further investigated, in addition to seasonal patterns, intraday and across other timescales. 

\section*{Acknowledgement}

This research was supported by the RiskEcon\textsuperscript{\textregistered}  Lab for Decision Metrics @ Courant Institute of Mathematical Sciences, NYU.

\bibliographystyle{plainnat}
\bibliography{paper}

\end{document}